\documentclass[12pt,preprint]{aastex}
\begin{document}
\title{ON THE UNIVERSALITY OF THE BOUND-ZONE PECULIAR VELOCITY PROFILE}
\author{Jounghun Lee}
\affil{Astronomy Program, Department of Physics and Astronomy, Seoul National University, 
Seoul 08826, Republic of Korea} 
\email{jounghun@astro.snu.ac.kr}
\begin{abstract}
A numerical evidence for the universality of the bound-zone peculiar velocity profile in a $\Lambda$CDM universe  is presented.  
Analyzing the dark matter halo catalogs from the Millennium-II simulation, we determine the average peculiar velocity profiles of 
the objects located in the bound zones around massive group-size halos at various redshifts and compare them to an analytic formula 
characterized by two parameters, the amplitude and slope of the profile.  
The best-fit values of the two parameters are found to be robust against the changes of the mass scales and the key cosmological parameters. 
It is also found that the amplitude and slope parameters of the bound-zone peculiar velocity profile are constant but only in the limited 
ranges of redshifts. 
In the dark matter dominated epoch corresponding to $z> 0.6$ the two parameters have constant values. 
In the transition period corresponding to $0.2\le z\le 0.6$ when the density of $\Lambda$ begins to exceed that of dark matter the two parameters 
grow almost linearly with redshifts. At later epochs $z<0.2$ when the $\Lambda$-domination prevails the two parameters regain constancy 
settling upon higher constant values. 
Noting that the length of the transition period depends on the 
amount of $\Lambda$ and speculating that the linear evolution of the profile with redshifts in the transition period is a unique feature of the 
$\Lambda$-dominated universe, we suggest that the redshift evolution of the bound-zone peculiar velocity profile should be a powerful local 
discriminator of dark energy candidates. 
\end{abstract}

\section{INTRODUCTION}\label{sec:intro}

In the field of cosmology, the turn-around radius of a bound object is rather a classical concept, defined as the  
radius that an overdense region of which the object has condensed out must have had at the moment when a perfect 
balance between the attraction of its self-gravity and the repulsion of the expanding spacetime was struck prior to 
the occurrence of its gravitational collapse \citep{GG72}. 
The very definition of the turn-around radius implies that it must be sensitively dependent upon the amount of dark matter, dark energy 
and the nature of gravity \citep[see, e.g.,][]{lahav-etal91,BS93,MB04,bor-etal12,fan-etal15}. 

Until recently, however, its importance and power as a cosmological discriminator has not been well recognized mostly 
because of the concern over the practical difficulty to measure it from observations. 
Unlike the viral radius of a bound object that is a nonlinear quantity and conventionally estimated as the radius at which the mass 
density reaches a few hundred times the critical density $\rho_{c}$ \citep[e.g.,][]{lahav-etal91}, the turn-around radius is a linear quantity 
and not readily estimated from the physical properties of a bound object observed in the present epoch. In principle, the turn-around radius 
of a massive object could be determined as the distance to the "zero velocity surface" \citep[e.g.,][]{kara-etal14} where the 
peculiar motion caused by the gravitational force of the given object cancels out the receding motion of the Hubble flow. 
But, the high degree of inaccuracy normally  involved in the measurements of the peculiar velocity field 
\citep[see][and references therein]{WF15} has so far made it unfavored to use the turn-around radius as a cosmological probe. 

Nevertheless, recently, \citet{PT14} recounted the advantages of using this classical concept as a test of the background 
cosmology. They first analytically proved the existence of a {\it time-independent direct} relation between the upper-limit of the 
turn-around radius and the density of dark energy, emphasizing the utmost utility of this relation as an invalidator of dark energy 
candidates including the standard cosmological constant $\Lambda$ \citep[see also][]{pavlidou-etal14}.
The other discriminative merit of the turn-around radius that \citet{PT14} pointed out is that it is not affected by complicated 
non-gravitational baryonic processes, which often obscure the powers of the standard diagnostics such as the redshift 
distortion effect, number counts of galaxy clusters, weak lensing effect and so on 
\citep[see, e.g.,][]{jing-etal06,rudd-etal08,BP13,hel-etal16}. 

In the wake of \citet{PT14}, probing attention has been paid to the turn-around radius and its prospect for an independent 
test of dark energy and gravity. For instance, \citet{faraoni16} have studied how the turn-around radius evolves in models with 
modified gravity (MG) and showed that a broad class of MG models could be ruled out through the measurements of the 
turn-around radii given its strong variation with the degree of MG. \citet{tan-etal16} studied analytically the evolution of the 
turn-around radius of a massive halo in a $\Lambda$CDM universe  and found its sensitivity on the amount of dark energy.
\citet{lee-etal15} enhanced the feasibility of employing the turn-around radius as a cosmological diagnostics by putting forth a 
novel methodology to measure it in practice. Their methodology is based on the numerical finding of \citet{falco-etal14} that 
the radial profile of the peculiar velocity field, $v_{p}(r)$, in the bound-zone around a massive object with virial mass $M_{\rm v}$ 
and radius $r_{\rm v}$ is well described by the following universal formula:  
\begin{equation}
\label{eqn:vprofile}
v_{p}(r) = - a\left(\frac{GM_{\rm v}}{r_{\rm v}}\right)^{1/2}\left(\frac{r_{\rm v}}{r}\right)^{b}\, ,
\end{equation}
where $a$ and $b$ are the amplitude and the slope parameters of the velocity profile, respectively, and $r$ is the 
radial distance from the center of the object. \citet{falco-etal14} demonstrated that in the bound zone corresponding to 
$3\le r/r_{v}\le 8$, Equation (\ref{eqn:vprofile}) with constant values of $a\approx 0.8\pm 0.2$ and $b\approx 0.42\pm 0.16$ 
agrees very well with the numerically obtained profile from a high-resolution $N$-body experiment, no matter 
how massive the bound objects are. 

Noting that at the turn-around radius (say, $r_{\rm t}$) Equation (\ref{eqn:vprofile}) should equal the velocity of the Hubble flow 
in magnitude, \citet{lee-etal15} suggested that the value of $r_{\rm t}$ be found as a solution to 
$a\left(GM_{\rm v}/r_{\rm v}\right)^{1/2}\left(r_{\rm v}/r_{\rm t}\right)^{b} = H_{0}r_{\rm t}$.
In other words, using the new methodology developed by \citet{lee-etal15} it is possible to estimate the turn-around radius of a massive 
object without having to measure the peculiar velocity field in the bound-zone as far as prior information on its viral mass is available.
\citet{lee-etal15} applied their methodology to a nearby isolated galaxy group, NGC5353/4 \citep{TT08,tully15}, and discovered that the 
estimated turn-around radius of the group exceeds the upper limit derived by \citet{PT14} for a $\Lambda$CDM cosmology.  
Yet, before interpreting their result as a tension against the $\Lambda$CDM ($\Lambda$+Cold Dark Matter) paradigm, \citet{lee-etal15} 
was emphatic about the need for further investigation of the validity of Equation (\ref{eqn:vprofile}) to which the success of their methodology 
was subject.  

Given that \citet{falco-etal14} derived the universal formula for $v_{p}(r)$ by measuring the peculiar velocities of {\it dark matter particles} 
located in the bound zone around the cluster-size halos, it should be important to examine whether or not the same formula also works 
for the case that $v_{p}(r)$ is determined from the peculiar velocity field of the bound-zone halos rather than that of dark matter particles 
since the latter is impossible to measure in practice.
Moreover, it should be also quite necessary to explore more extensively if and how the peculiar velocity profile of the bound-zone 
halos evolve with redshifts. Although \cite{falco-etal14} claimed that Equation (\ref{eqn:vprofile}) is universal with constant parameters, 
it may be true only for the case of the bound-zone dark matter particles but not for the case of the bound-zone halos, since 
it would take longer time for the halos to become stabilized in the bound-zone after their formation.  If the peculiar velocity profile of the 
bound-zone halos turns out not to be constant, then its evolution relation as a function of redshift might also exhibit a distinct 
dependence on the background cosmology. 
In this paper, we are to perform these follow-up tasks by measuring the peculiar velocity profile of the bound-zone objects around 
massive group-sized halos from the numerical data based on high-resolution N-body simulations. 

\section{PECULIAR VELOCITY PROFILES OF THE BOUND-ZONE OBJECTS}\label{sec:vpr}

We utilize the Friends-of-Friends (FoF) halo catalogs produced by the Millennium-II simulations for a $\Lambda$CDM 
cosmology with $\Omega_{m}=0.25,\ \Omega_{\Lambda}=0.75,\ \Omega_{b}=0.045,\ h=0.73,\ \sigma_{8}=0.9,\ n_{s}=1$ 
\citep{millennium2}.  
The simulations were performed with high resolution in a periodic box of volume $100^{3}\,h^{-3}$Mpc$^{3}$ with $2160^{3}$ DM 
particles each of which has equal mass of $6.885\times 10^{6}\,h^{-1}M_{\odot}$. The Millennium-II simulations identified the DM halos 
by applying the FoF finder to each snapshot recorded at various redshifts.  To have a full description of the Millennium-II N-body 
simulation and the FoF halo catalogs, refer to \citet{lemson-etal06} and \citet{millennium2}.

Various physical properties of each DM halo such as its mass ($M_{200}$),  radius ($r_{200}$), number of its 
component DM particles, positions $({\bf x})$ and peculiar velocities $({\bf v}_{p})$ are provided in 
the Millennium-II halo catalog.  Here $r_{200}$ is the radius at which the mean mass density $\rho$ equals $200$ times the critical mass 
density $\rho_{\rm c}$ and $M_{200}$ is the mass contained within a sphere with radius $r_{200}$, satisfying 
$M_{200} = (4\pi/3)(200\rho_{\rm c})r^{3}_{200}$. For our analysis throughout this paper, we regard $r_{200}$ and $M_{200}$ as the 
viral radius  and viral mass, respectively. 

We first make a sample of the group-sized halos by selecting those halos whose viral masses, $M_{\rm g}$, exceed 
$10^{13}\,h^{-1}M_{\odot}$ from the Millennium-II halo catalogs. For each selected group-sized halo, we search for the bound zone 
halos which satisfy the following two conditions: First, the virial mass $M_{\rm b}$ of the bound-zone halos exceeds a certain threshold 
value, $M_{\rm b,c}$. To begin with, we set $M_{\rm b, c}$ at $10^{8}\,h^{-1}M_{\odot}$. 
Second, the radial distance, $r$, from the group center is in the range of $3\le r/r_{\rm v}\le 8$ where $r_{\rm v}$ is the viral radius of the 
group \citep{falco-etal14}. This range corresponds to the bound-zone where the peculiar velocity field can be treated as a 
linear quantity. The inner range $r/r_{\rm v}< 3$ is called the "infall zone" where the peculiar velocity field behaves nonlinearly, while the 
outer range $r/r_{\rm v}> 8$ is the "Hubble zone" where the peculiar velocity field is effectively zero \citep{zu-etal14, lee-etal15}. 
In our analysis, we focus only on the bound-zone peculiar velocity field. 

Projecting the peculiar velocities of the bound-zone halos, ${\bf v}_{p}$, onto the radial direction toward the group center, we 
investigate how the rescaled  value $\tilde{v}_{p}\equiv v_{p}/V_{c}$ with $V_{c}\equiv \left[(GM_{\rm v})/r_{\rm v}\right]^{1/2}$ 
changes as a function of the rescaled radial distance $\tilde{r}\equiv r/r_{v}$.  
From here on, for simplicity, a group-sized halo in our sample ($M_{\rm g}\ge 10^{13}\,h^{-1}M_{\odot}$ ) is called a group and the magnitude 
of the component of the peculiar velocity of a bound-zone halo along the radial direction is called the bound-zone peculiar velocity, denoted by 
$v_{\rm p}(r)$.  Note that the massive cluster-sized halos (say, $M_{\rm g}\ge 10^{14}\,h^{-1}M_{\odot}$ ) are included in our sample of the 
groups. 

\begin{figure}
\begin{center}
\plotone{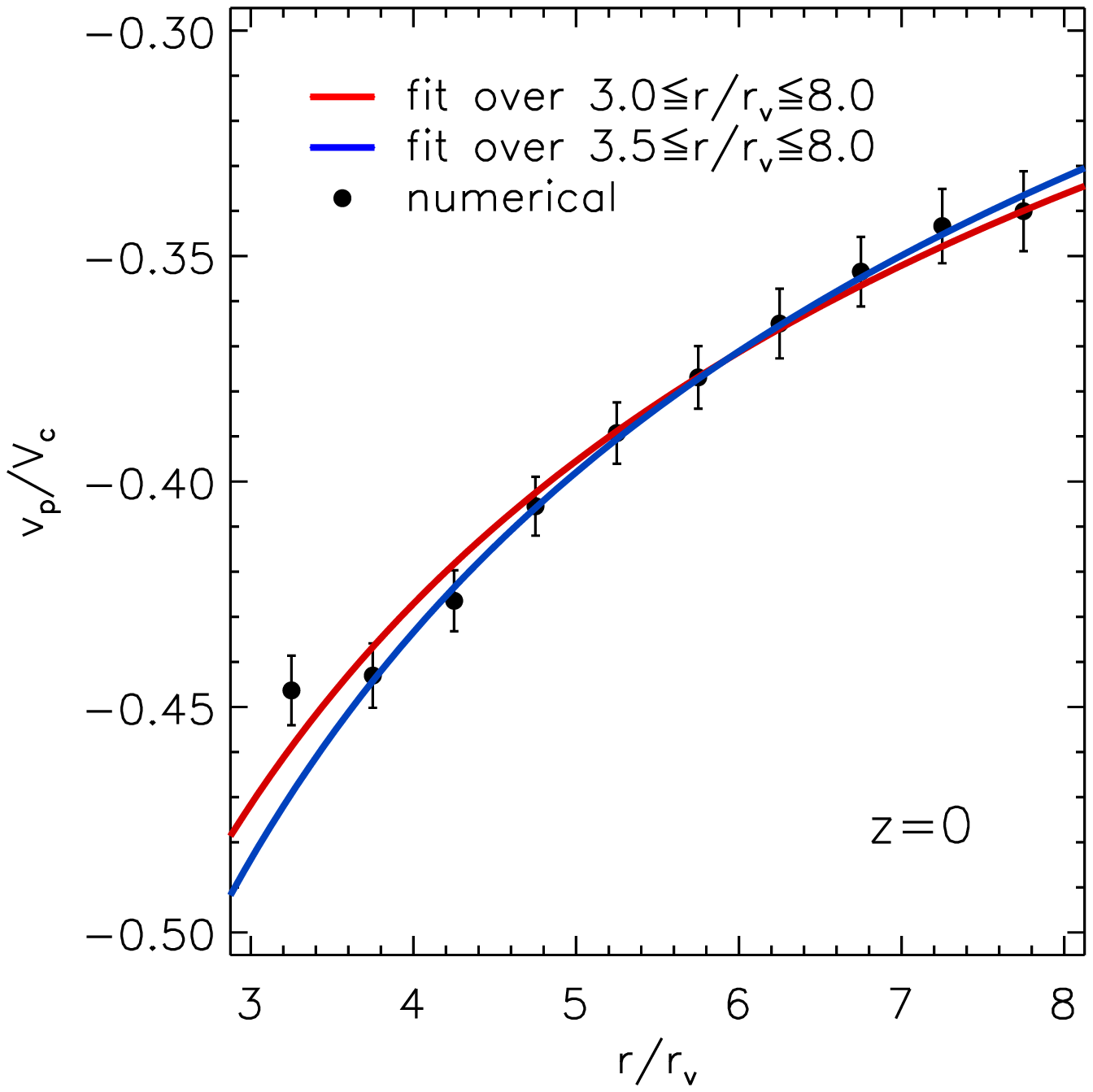}
\caption{Peculiar velocity profile of the bound-zone halos with virial masses $M_{\rm b}\ge 10^{8}\,h^{-1}\,M_{\odot}$ 
around the groups with viral mass $M_{\rm g}\ge 10^{13}\,h^{-1}\,M_{\odot}$ at $z=0$. The black filled circles with 
error bars  are the numerical results from the Millennium II simulations, while the red and the 
blue solid lines represent the best-fit models over the range of $3\le r/r_{\rm v}\le 8$ and 
$3.5\le r/r_{\rm v}\le 8$, respectively.}
\label{fig:vpr_z0}
\end{center}
\end{figure}

For each group, we bin the rescaled radial distance $\tilde{r}$ and count the numbers of the bound-zone halos, $N_{\rm b}$, whose 
radial distances belong to each bin. Let ${\tilde v}_{p}^{\alpha, \beta}$ denote the rescaled peculiar velocity of the $\alpha$-th 
bound-zone halo around the $\beta$-th group at a given bin, $\tilde{r}$. 
The peculiar velocity profile of the bound-zone halos averaged over a total of $N_{\rm g}$ groups in our sample is evaluated as 
\begin{equation}
\tilde{v}_{p}(\tilde{r}) = \frac{1}{N_{\rm g}}\sum_{\beta}\left[\frac{1}{N_{\rm b}}\sum_{\alpha}\tilde{v}_{p}^{\alpha,\beta}(\tilde{r})\right]\, .
\end{equation}
Figure \ref{fig:vpr_z0} plots the resulting bound-zone peculiar velocity profile at $z=0$ from the Millennium-II simulations as filled circular dots. 
Here, the errors $\sigma_{\rm v}$ are computed as the standard deviation in the measurements of the average value of 
$\tilde{v}_{p}(\tilde{r})$. 

Now that the numerical result for $\tilde{v}_{p}(\tilde{r})$ has been obtained, we fit it to the analytic model, 
Equation (\ref{eqn:vprofile}), by adjusting the amplitude and slope parameters, $a$ and $b$, with the help of the maximum likelihood 
method. Basically, we search for $(a, b)$ which maximizes the joint probability, 
$p(a,b) \propto \exp\left[-\chi^{2}(a,b)/2\right]$,  where the {\it reduced} chi-square $\chi^{2}(a,b)$ is given as 
\begin{equation}
\label{eqn:chi2}
\chi^{2}_{\nu}(a,b) = \frac{1}{\left(N_{\rm bin}-2\right)}\sum_{i=1}^{N_{\rm bin}} 
\frac{\left[\tilde{v}_{p}(\tilde{r}_{i})- \tilde{v}^{2}_{\rm p, model}(\tilde{r}_{i};a,b)\right]^{2}}{\sigma^{2}_{\rm v}(\tilde{r}_{i})}\, , 
\end{equation}
where $N_{\rm bin}$ is the number of the bins of the rescaled radial distance $\tilde{r}$, $\tilde{v}_{p}(\tilde{r}_{i})$ denotes the value of the 
numerically obtained peculiar velocity profile of the bound zone halos at the $i$-th bin $\tilde{r}_{i}$, while  $\tilde{v}_{\rm p,model}(\tilde{r}_{i};a,b)$ 
represents the analytic prediction of Equation (\ref{eqn:vprofile}) at $\tilde{r}_{i}$ where the amplitude and slope parameters have the values of 
$a$ and $b$, respectively. Since the analytic model has two adjustable parameters, the degree of freedom for the chi-square equals 
$N_{\rm bin}-2$ \citep{WJ12}.

In Figure \ref{fig:vpr_z0} the fitting model with the best parameter values of $a$ and $b$ is shown as red solid line. Although the overall 
accord between the numerical and the fitting results is quite good, the numerical value, $\tilde{v}_{p}(\tilde{r}_{i})$ at the left-most 
bin appears to depart from the overall trend. That is, the numerical result shows a bit flatter shape than the analytic model in the lowest 
$\tilde{r}$ bin. We suspect that it is because in the boundary, $\tilde{r}\approx 3$, between the infall and the bound zones, the nonlinear 
aspect of the gravitational force of a group begins to be effective, leading to a departure of the peculiar velocity from the linear behavior 
at $r\approx 3r_{\rm v}$. 
To remove this contamination caused by the nonlinear effect from the construction of the bound-zone peculiar velocity profile, we exclude the 
left-most bin from the procedure and reperform the fitting over the narrower range of $3.5\le \tilde{r}\le 8$, the result of which is shown as 
blue solid line in Figure \ref{fig:vpr_z0}. As can be seen, the narrower range yields a better fit to the numerical result. The excellent match 
between the numerically obtained profile and Equation (\ref{eqn:vprofile}) confirms the validity of the analytic formula of \citet{falco-etal14} 
even for the case that the bound-zone peculiar velocity profile is constructed not from DM particles but from the halos.

\begin{figure}
\begin{center}
\plotone{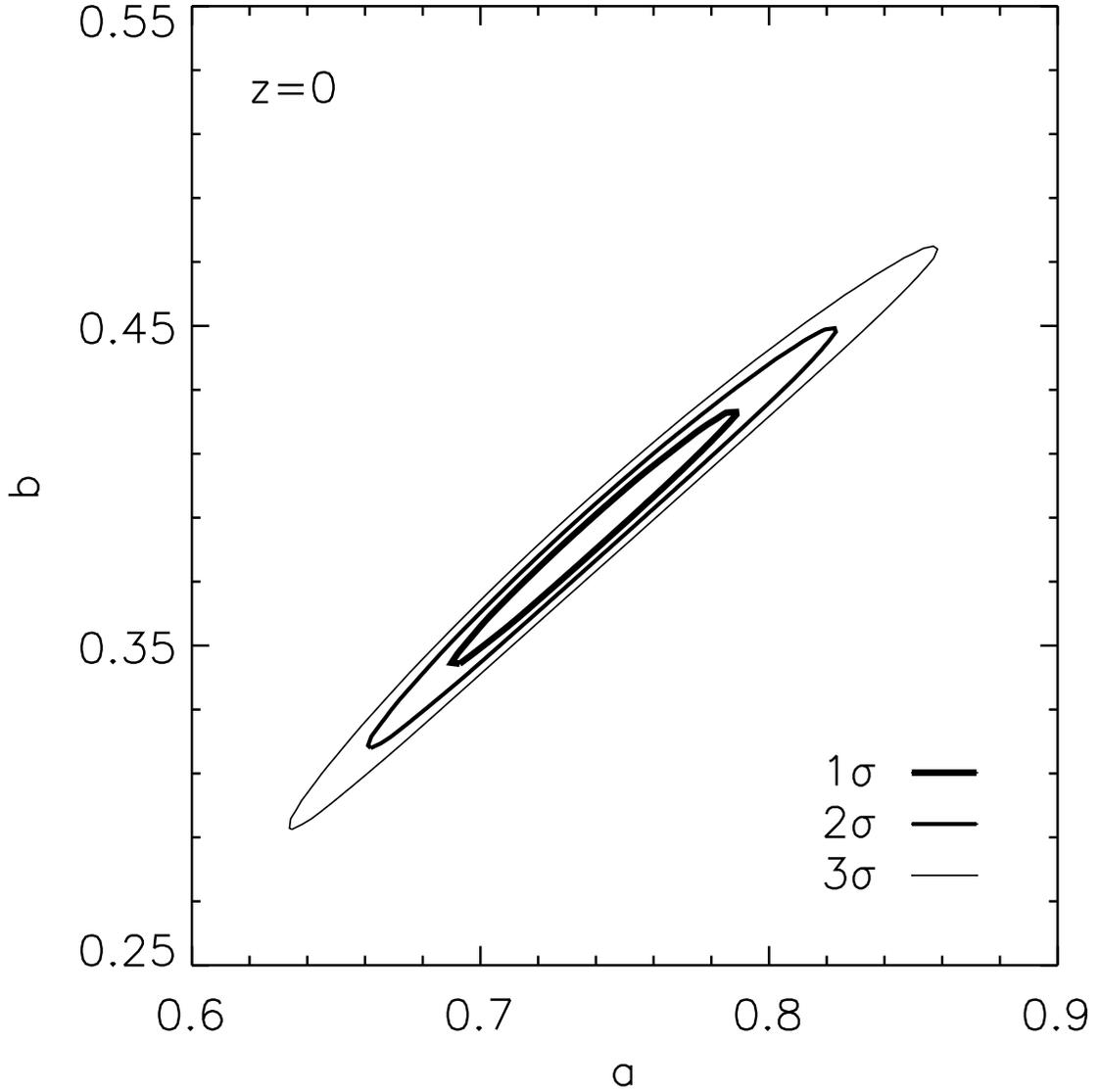}
\caption{$1\sigma$, $2\sigma$ and $3\sigma$ contours of $\chi^{2}$ 
(thick, thin and the thinnest solid lines, respectively) in the $a$-$b$ plane. }
\label{fig:cont_z0}
\end{center}
\end{figure}

\begin{figure}
\begin{center}
\plotone{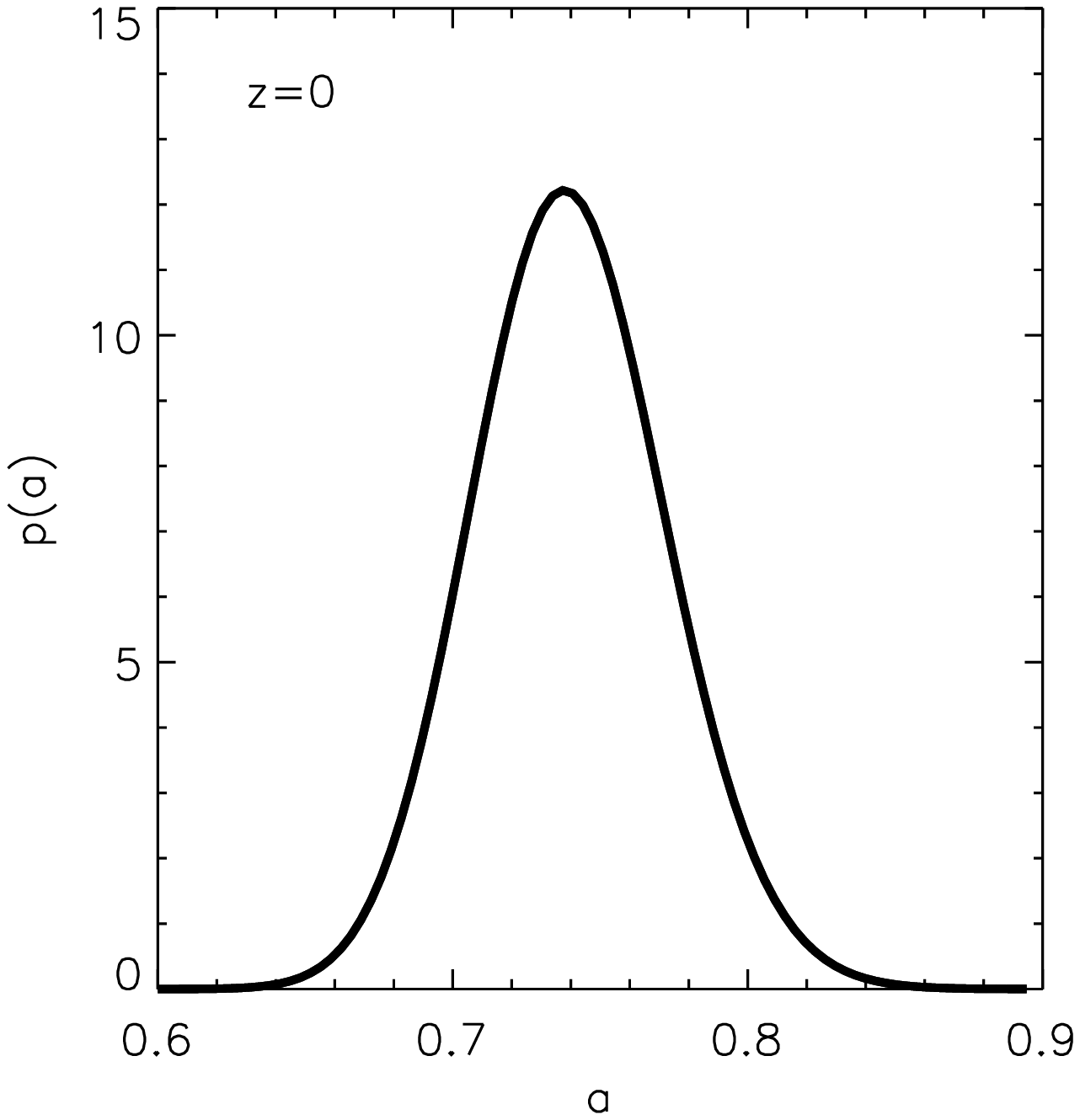}
\caption{Probability density distribution of the amplitude parameter $a$ obtained by 
marginalizing the joint probability density distribution over the slope parameter $b$ at $z=0$.}
\label{fig:proa}
\end{center}
\end{figure}

\begin{figure}
\begin{center}
\plotone{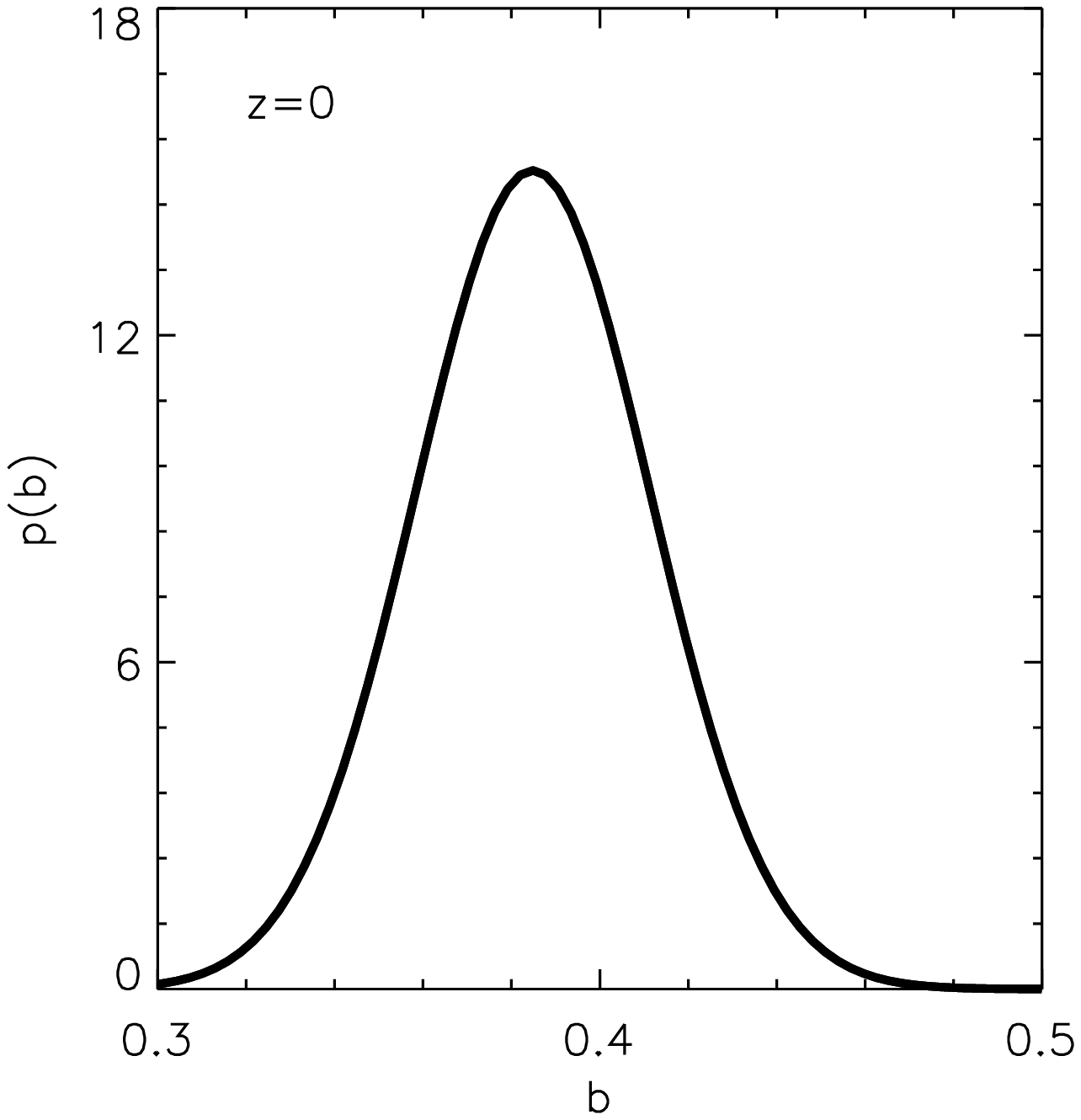}
\caption{Probability density distribution of the slope parameter $b$ obtained by 
marginalizing the joint probability density distribution over the amplitude parameter $a$ at $z=0$.}
\label{fig:prob}
\end{center}
\end{figure}

Figure \ref{fig:cont_z0} plots the $1\sigma,\ 2\sigma,\ 3\sigma$ contours of $p(a,b)$ at $z=0$ in the $a$-$b$ plane as the thickest, thick and thin 
solid lines, respectively, under the assumption that the joint probability density distribution $p(a,b)$ is well approximated as a multi-variate
Gaussian distribution. Here, the $1\sigma,\ 2\sigma$ and $3\sigma$ contours show the areas in the $a$-$b$ plane over which the probability, 
$\int\,da\,\int\,db\,p(a,b)$, has the values of $0.68,\ 0.95$ and $0.99$, respectively. 
To compute the errors associated with the measurements of the best-fit values of $a$ and $b$, we first marginalize the joint probability density 
$p(a,b)$ to evaluate the probability density distributions of the two parameters, $p(a)$ and $p(b)$, as $p(a)=\int\,db\,p(a,b)$ 
and $p(b)=\int\,da\,p(a,b)$, the results of which are displayed as black solid lines in Figures \ref{fig:proa} and \ref{fig:prob}, respectively.  
As can be seen, both of $p(a)$ and $p(b)$ are nearly symmetric about the values at which the probability densities reach the highest values.  
The marginalized errors associated with the measurements of the best-fit parameters are calculated as 
$\int_{a_{\rm max}}^{a_{\rm max}+\sigma_{a+}}\,da\,p(a) = 0.34,\ \int_{a_{\rm max}-\sigma_{a-}}^{a_{\rm max}}\,da\,p(a) = 0.34,\ 
\int_{b_{\rm max}}^{b_{\rm max}+\sigma_{b+}}\,db\,p(b) = 0.34,\ \int_{b_{\rm max}-\sigma_{b-}}^{b_{\rm max}}\,db\,p(b) = 0.34$, 
where $a_{\rm max}$ and $b_{\rm max}$ represent the most probable values of $a$ and $b$.

The first row of Table \ref{tab:ab_z} lists the number of the groups ($N_{\rm g}$), the total number of the bound-zone halos 
($N_{\rm T, b}$), the best-fit values of the amplitude and slope parameters ($a$ and $b$), and the minimum value of the reduced 
chi-square, $\chi_{\nu}^{2}\equiv \chi^{2}/(N_{\rm bin}-2)$ at $z=0$.  Note that the best-fit ranges of the two parameters, $a=0.74\pm 0.03$ and 
$b=0.38\pm 0.03$ found in the current work are well overlapped with those found in the original work of \citet{falco-etal14}, $a=0.8\pm 0.2$ and 
$b=0.42\pm 0.16$.  
This result implies that the bound-zone peculiar velocity profile constructed from the bound-zone halos can be described by the same analytic 
formula, Equation (\ref{eqn:vprofile}) with the same parameters , that \citet{falco-etal14} originally derived from the bound-zone DM particles.  
As mentioned in Section \ref{sec:intro}, the bound-zone peculiar velocity profile can be reconstructed only from the bound objects in real observations,  
this result holds up the reliability and practicality of the analytic formula of \citet{falco-etal14}. 

Recall that for the numerical simulations utilized by \citet{falco-etal14} the normalization amplitude of the initial power spectrum was set at 
$\sigma_{8}=0.76$ \citep{wmap3}, substantially different from the value of $\sigma_{8}=0.9$ adopted for the Millennium-II simulations \citep{millennium2}. 
Recall also that in the original work of \citet{falco-etal14} the ratio of the mass overdensity within the virial radius to $\rho_{\rm c}$ was set at 
$\Delta = 97.8$ while in the current work based on the Millennium-II simulations it is $\Delta = 200$. In spite of these differences, the best-fit values 
of the amplitude and slope of the bound-zone peculiar velocity profile obtained in our fitting procedure agree quite well with the original values determined 
\citet{falco-etal14}. This result implies that the analytic formula of the bound-zone peculiar velocity profile with constant amplitude and slope should be 
insensitive to the key cosmological parameter values as well as to the details of the way in which the virial radius/mass is defined.
However, before concluding that the bound-zone peculiar velocity profile is truly universal, a further investigation should be performed about the 
constancy of its amplitude and slope against the change of redshifts.

\section{EVOLUTION OF THE BOUND-ZONE PECULIAR VELOCITY PROFILES}\label{sec:vpr_z}

\begin{figure}
\begin{center}
\plotone{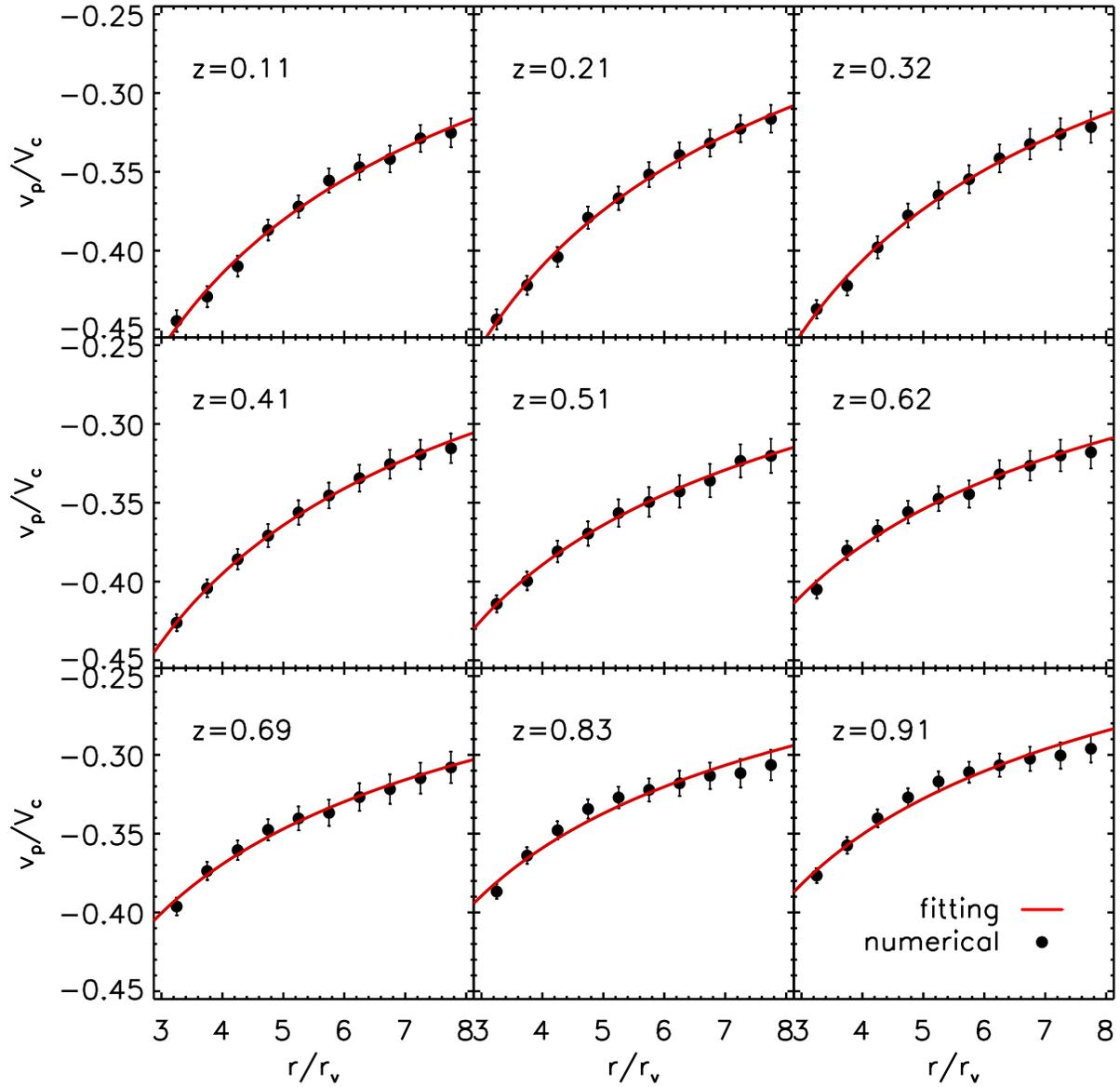}
\caption{Peculiar velocity profiles of the bound-zone halos at nine different redshifts. 
In each panel the filled black circles with errors and the red solid lines represent the numerical 
results and the best-fit model, respectively.} 
\label{fig:vpr_z}
\end{center}
\end{figure}

We refollow the whole procedures arranged in Section \ref{sec:vpr} with the halos identified at higher redshifts: 
selecting the groups with virial mass $M_{\rm g}\ge 10^{13}\,h^{-1}\,M_{\odot}$;  searching for the bound-zone halos at 
$3\le \tilde{r}\le 8$ with virial mass $M_{\rm b}\ge 10^{8}\,h^{-1}\,M_{\odot}$; 
constructing the peculiar velocity profile of the bound-zone halos $\tilde{v}_{p}$ as a function of $\tilde{r}$; fitting $\tilde{v}_{p}$ to 
Equation (\ref{eqn:vprofile}) and determine the best-fit parameters, $(a, b)$, that minimize $\chi^{2}$ given in 
Equation (\ref{eqn:chi2}) or equivalently maximizes  $p(-\chi^{2}/2)$. 
Figure \ref{fig:vpr_z} shows the numerical results (filled circular dots with errors) along with the best-fit models (red solid lines) 
at nine different redshifts in the range of $0.1\le z< 1$.  It is obvious that the fitting formula, Equation (\ref{eqn:vprofile}), is 
indeed as successful at higher redshifts as at the present epoch in matching the numerical results. Unlike the case of $z=0$, 
we do not find a departure of the numerical data point from the overall trend in the left-most bin, $\tilde{r}=3$, at $z\ge 0.1$ and 
thus use the whole range of $3\le \tilde{r}\le 8$ without excluding the left-most bin. This phenomenon indicates that the nonlinear 
effect on the peculiar velocity field in the boundary ($\tilde{r}=3$) between the infall and the bound zones becomes significant only 
at the present epoch. 

\begin{figure}
\begin{center}
\plotone{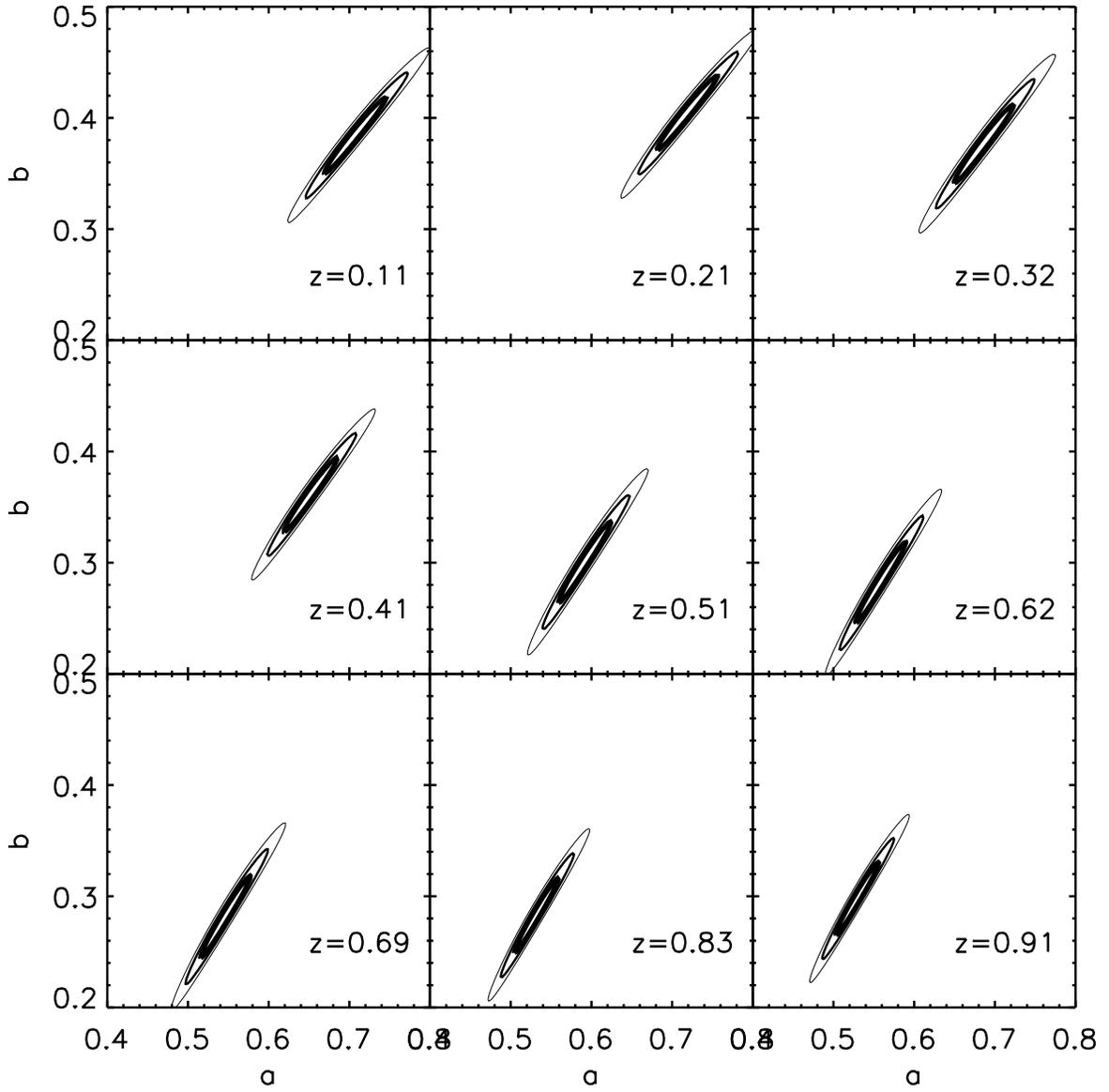}
\caption{Same as Figure \ref{fig:cont_z} but at nine higher redshifts.}
\label{fig:cont_z}
\end{center}
\end{figure}

\begin{figure}
\begin{center}
\plotone{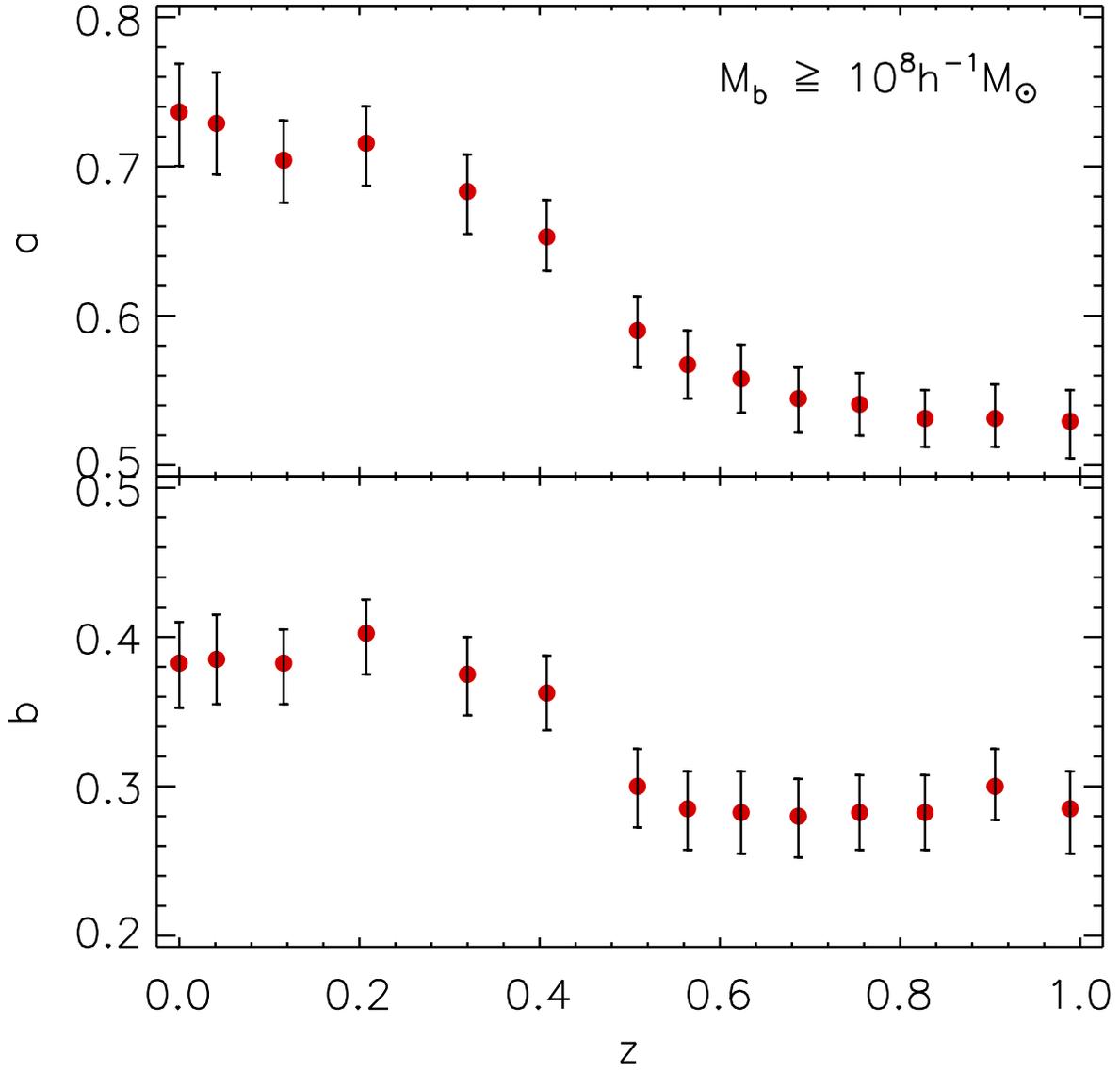}
\caption{Variation of the best-fit parameters, $a$ and $b$, in Equation (\ref{eqn:vprofile}) 
with redshifts in the top and bottom panels.}
\label{fig:ab_z}
\end{center}
\end{figure}

Figure \ref{fig:cont_z} show the same as Figure \ref{fig:cont_z0} but at nine different epochs. As can be seen, at earlier epochs the peculiar 
velocity profiles of the bound-zone halos are characterized by lower amplitudes and lower slopes.
To see more explicitly how the amplitude and slope of the peculiar velocity profile of the bound-zone halos change with 
redshift, we plot the best-fit values of $a$ and $b$ versus $z$ in Figure \ref{fig:ab_z}. In the high-$z$ section ($z>0.6$), 
the best-fit amplitude and slope appear to be uniform, independent of $z$. 
In the intermediate-$z$ section ($0.2\le z\le 0.6$), the best-fit parameters vary with $z$, making 
a gradual transition toward higher values.  In the low-$z$ section ($z< 0.2$), the two parameters reexhibit uniformity, 
being constant but at higher values than those in the high-$z$ section ($z>0.6$). 

Noting in Figure \ref{fig:cont_z} that the transitions of $a$ and $b$ to higher values start at the epoch when $\Lambda$ begins 
to dominate the energy density of the universe, we ascribe the redshift-evolution of the bound-zone peculiar velocity profile to 
the effect of $\Lambda$.  
The higher slope of the bound-zone peculiar velocity profile in the low-$z$ section can be understood as follows.
The stronger repulsive force of the accelerating space-time after the onset of $\Lambda$ would lead the peculiar velocities of the 
bound-zone halos to decrease more rapidly in magnitude with the radial distance. To understand the higher amplitude 
of the bound-zone peculiar velocity profile at $z\le 0.6$, one has to recall that the amplitude, $a$, is proportional to the linearly 
extrapolated value of $v_{p}(r)$ to the viral radius, $r_{\rm v}$.  For the gravitational force to resist the repulsive force
of the accelerating space-time at the viral radius in the $\Lambda$-dominated epoch, it has to have a higher value, which 
yields higher amplitude of $v_{p}(r_{\rm v})$.

\begin{figure}
\begin{center}
\plotone{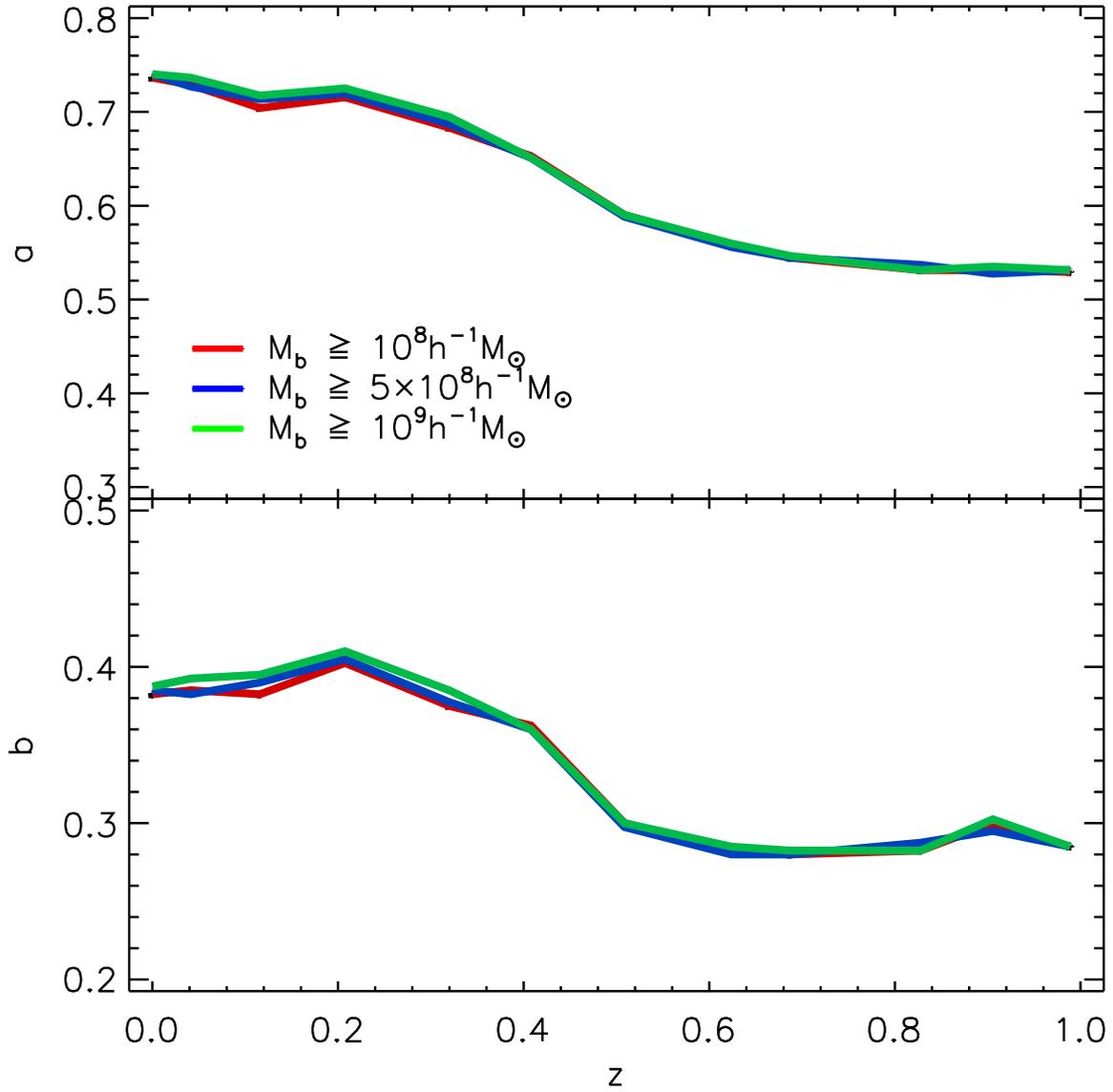}
\caption{Same as Figure \ref{fig:ab_z} but for three different cases of the cuts imposed on the virial masses 
of the bound-zone halos.}
\label{fig:abz_mb}
\end{center}
\end{figure}

To see if this trend is robust against the change of the mass cut-off scale of the bound-zone halos, we impose a higher mass cut 
on the virial masses $M_{\rm b}$ of the bound-zone halos and repeat the whole process. 
Figure \ref{fig:abz_mb} shows the evolution of the amplitude and slope of the bound-zone peculiar velocity profiles for three different cases 
of the cuts imposed in the masses of the bound-zone halos.  As can be seen, the three cases show the same trend. The amplitude and 
slope of the bound-zone peculiar velocity profile are constant when either DM or $\Lambda$ dominates but evolving linearly with $z$ 
when the energy density of $\Lambda$ is comparable to that of DM.  
Despite that there is an order of magnitude change in the mass cut-off scales of the bound-zone halos, there is no substantial change in 
the best-fit values of $a$ and $b$ in the whole redshift range of $0\le z\le 1$, which implies that the analytic formula of the bound-zone peculiar 
velocity profile is indeed robust against the variation of the mass cut-off scale of the bound-zone halos.  Given that what can be measured 
accurately from observations is usually the properties of the luminous massive galaxies but not those of the low-mass dwarf galaxies, 
our results offers a good prospect for the practical application of the bound-zone peculiar velocity profile in real observations. 

\section{SUMMARY AND DISCUSSION}\label{sec:con}

This work is a follow-up to \citet{lee-etal15} who proposed a new methodology to estimate the turn-around radius of a bound object without 
resorting to the inaccurate measurements of the peculiar velocity field.  Applying their methodology to the bound-zone galaxies of the 
NGC 5353/4 group, \citet{lee-etal15} have found that the turn-around radius of the group exceeds the upper limit set by the standard 
$\Lambda$CDM cosmology.  
Since the success of the methodology of \citet{lee-etal15} was contingent upon the validity of the analytic formula derived by 
\citet{falco-etal14} for the bound-zone peculiar velocity profile, it was required to test extensively the robustness of the formula before 
interpreting the estimated turn-around radius of the NGC 5353/4 group as a counter-evidence against the $\Lambda$CDM cosmology.
In the current work, we have performed the required test by utilizing the high-resolution Millennium-II simulations \citep{millennium2}.
First, we have examined if the same analytic formula is also applicable to the case that the peculiar velocities were measured not from 
individual DM particles but from the bound-zone galaxies or halos.  We have also investigated if the bound-zone peculiar velocity profile 
is truely universal, being well described by the same redshift-independent formula, as claimed by \citet{falco-etal14}. 

Analyzing the halo catalogs in the redshift range of $0\le z\le 1$ produced by the Millennium-II simulations \citep{millennium2}, we have 
constructed the peculiar velocity profiles of the bound-zone halos with mass $M_{\rm b}\ge 10^{8}\,h^{-1}\,M_{\odot}$ around massive 
group-size halos with mass $M_{\rm g}\ge 10^{13}\,h^{-1}\,M_{\odot}$ and determined the amplitude and slope of the bound-zone peculiar velocity 
profile by comparing the formula to the numerical results. We have shown that the peculiar velocity profile of the bound-zone halos 
can be well approximated by the same analytic model of \citet{falco-etal14}, even when the profile is constructed not from dark matter 
but from the halos. However, it has been found that the amplitude and slope parameters, $a$ and $b$, of the bound-zone peculiar velocity 
profile do not show perfect constancy  but evolve with redshifts:  In the high redshift range $0.6< z\le 1$, the two parameters are found to have 
constant values around $a=0.53$ and $b=0.28$. In the intermediate redshift range $0.2\le z\le 0.6$, both of the two parameters increase almost 
linearly with the decrement of $z$. In the low redshift range $z\le 0.2$ the constancy of the two parameters is recovered and the best-fit values of 
$a$ and $b$ are found to agree fairly well with the original values of $a=0.42$ and $b=0.8$ determined by \citet{falco-etal14}. 

Noting the coincidence between the epoch when the redshift evolution of the two parameters starts and the epoch when the density of 
$\Lambda$, $\rho_{\Lambda}$, begins to exceed that of dark matter, $\rho_{\rm m}$,  in the universe, we have claimed that the growing effect 
of $\Lambda$ should be responsible for the variation of the amplitude and slope of the bound-zone peculiar velocity profile in the intermediate 
redshift range $0.2\le z\le 0.6$.  We have explained that the accelerating space-time in the $\Lambda$-dominated era has an effect of 
steepening the peculiar velocity profile and reducing the circular velocity by defeating more drastically the gravitational attraction of 
massive halos with the increment of the radial distances. 

We have also verified that the bound-zone peculiar velocity profile is robust against the variation of the initial conditions of the $\Lambda$CDM 
universe, demonstrating that in spite of the different values of the key cosmological parameters adopted by the Millennium-II simulations from 
that of \citet{falco-etal14} the best-fit values of the amplitude and slope parameters of the bound-zone peculiar velocity profile determined by 
our analysis agree well the original values found by \citet{falco-etal14}. Along with our another finding that the bound-zone peculiar velocity profile 
is also robust against the variation of the mass cut-off scale of the bound-zone objects, our results altogether lead us to the bottom line that in the 
low-redshift range $z\le 0.2$ when the $\Lambda$-domination prevails the bound-zone peculiar velocity profile is indeed universal, providing a 
compelling numerical evidence for the original claim of \citet{falco-etal14}. 

We speculate that the length of the transition period when the amplitude and slope parameters of the bound-zone peculiar velocity profile 
grow linearly with the decrement of redshifts should be a powerful estimator of the amount of $\Lambda$ in the {\it local} universe, since it depends 
sensitively when the $\Lambda$-domination starts and prevails.  
 We also speculate that the linear growth of the amplitude and slope parameters of the bound-zone peculiar velocity profile during the transition 
period of $0.2\le z\le 0.6$ may be a unique signature of the $\Lambda$CDM cosmology: In alternative dark energy models or in MG models,  the 
growth rate would deviate from being constant and might assume a dependence on the redshift, given that it reflects how rapidly the acceleration 
of spacetime cracks the balance between the gravity and the Hubble expansion.  It might be possible to distinguish between 
alternative cosmologies by measuring the growth rate of the amplitude and slope of the bound-zone peculiar velocity profile during the transition 
period from observations. Of course, what should precede is to investigate quantitatively at what rates the characteristic parameters of the 
bound-zone peculiar velocity profile grow in alternative models. Our future work is in this direction. 

\acknowledgements
We thank an anonymous referee for very helpful comments.
This work was supported by a research grant from the National Research Foundation (NRF) of Korea to the Center for 
Galaxy Evolution Research  (NO. 2010-0027910) and partially by the support of the Basic Science 
Research Program through the NRF of Korea funded by the Ministry of Education (NO. 2013004372).
The Millennium-II Simulation databases used in this paper and the web application providing online access to them were constructed 
as part of the activities of the German Astrophysical Virtual Observatory (GAVO).

\clearpage
\clearpage

\clearpage
\clearpage
\clearpage
\clearpage
\clearpage
\clearpage

\clearpage
\begin{deluxetable}{cccccc}
\tablewidth{0pt}
\setlength{\tabcolsep}{5mm}
\tablecaption{Redshifts, number of the groups, number of the bound-zone halos around 
the groups, best-fit values of the amplitude and slope parameters of the bound-zone peculiar velocity 
profile with marginalized errors, and the reduced chi-square value for the case of 
$M_{\rm b}\ge 10^{8}\,h^{-1}\,M_{\odot}$.}
\tablehead{z & $N_{\rm g}$ & $N_{\rm T,b}$ & $a$ & $b$ & $\chi^{2}_{\nu}$}
\startdata
$0$  & $301$ & $604069$&$0.73\pm 0.03$ & $0.38\pm 0.03$ & $0.11$\\
$0.11$ & $297$ & $674335$&$0.70\pm 0.03$ & $0.38\pm 0.02$ & $0.28$\\
$0.21$ & $282$ & $691961$&$0.72\pm 0.02$ & $0.40\pm 0.02$ & $0.13$\\
$0.32$  & $279$ & $751384$&$0.68\pm 0.03$ & $0.38\pm 0.03$ & $0.20$\\
$0.41$ & $271$ & $797699$&$0.65\pm 0.02$ & $0.36\pm 0.03$ & $0.05$\\
$0.51$ & $279$ & $838630$&$0.60\pm 0.02$ & $0.30\pm 0.03$ & $0.06$\\
$0.62$  & $257$ & $828457$&$0.56\pm 0.02$ & $0.28\pm 0.03$ & $0.28$\\
$0.69$ & $248$ & $813384$&$0.54\pm 0.02$ & $0.28\pm 0.03$ & $0.23$\\
$0.83$ & $222$ & $750696$&$0.53\pm 0.02$ & $0.28\pm 0.03$ & $0.83$\\
$0.91$ & $208$ & $689159$&$0.53\pm 0.02$ & $0.30\pm 0.03$ & $0.64$\\
$0.99$ & $203$ & $691205$&$0.53\pm 0.02$ & $0.28\pm 0.03$ & $0.33$
\enddata
\label{tab:ab_z}
\end{deluxetable}

\end{document}